\documentclass[%
 aip,
 amsmath,amssymb,
 reprint,%
]{revtex4-1}

\usepackage{graphicx}
\usepackage{dcolumn}
\usepackage{bm}

\usepackage[utf8]{inputenc}
\usepackage[T1]{fontenc}
\usepackage{mathptmx}
\DeclareUnicodeCharacter{2212}{-}
\DeclareUnicodeCharacter{2264}{$\leq$}
\DeclareUnicodeCharacter{0327}{ą}
\begin{document}

\preprint{AIP/123-QED}

\title[Sample Title]{Thin films of the $\alpha$-quartz $Si_xGe_{1-x}O_2$ solid-solution}

\author{Silang Zhou}
\author{Jordi Antoja-Lleonart}%
 
\author{Václav Ocelík}

\author{Beatriz Noheda}
\email{b.noheda@rug.nl}
\affiliation{%
 Zernike Institute for Advanced Materials, University of Groningen, Nijenborgh 4, 9747AG Groningen, The Netherlands
}%

\date{\today}

\begin{abstract}
$SiO_2$ with the $\alpha$-quartz structure is one of the most popular piezoelectrics. It is widely used in crystal oscillators, bulk acoustic wave (BAW) devices, surface acoustic wave (SAW) devices, and so on. $GeO_2$ can also be crystallized into the $\alpha$-quartz structure and it has better piezoelectric properties, with higher piezoelectric coefficient and electromechanical coupling coefficients, than $SiO_2$. Experiments on bulk crystals and theoretical studies have shown that these properties can be tuned by varying the Si/Ge ratio in the $Si_xGe_{1-x}O_2$  solid solution. However, to the best of our knowledge, thin films of $Si_xGe_{1-x}O_2$ quartz have never been reported. Here we present the successful crystallization of $Si_xGe_{1-x}O_2$ thin films in the $\alpha$-quartz phase on quartz substrates ($SiO_2$) with x up to 0.75. Generally, the films grow semi-epitaxially, with the same orientation as the substrates. Interestingly, the $Si_{0.75}Ge_{0.25}O_2$ composition grows fully strained by the quartz substrates and this leads to the formation of circular quartz domains with an ordered Dauphin\'{e} twin structure. These studies represent a first step towards the optimization of piezoelectric quartz thin films for high frequency ($>$ 5 GHz) applications.
\end{abstract}

\maketitle

\section{\label{sec:level1}Introduction}

$\alpha$-quartz, $SiO_2$, crystals are currently used as clocks in basically all integrated circuits and as frequency stabilizers in radio transmitters and receivers. The availability of $\alpha$-quartz films with sub-micron thickness and quality factor similar to those of the single crystals, would enable the new generation of devices for information and communication technology beyond 5 GHz. Efforts in this direction are still in their infancy guided by seminal work on thin films of $SiO_2$ synthesised by chemical solution deposition methods \cite{Jolly2021SoftChemistryAssistedSystem,Carretero-Genevrier2013b,Zhang2019TailoringFilms,Zhang2020Micro/NanostructureSilicon}. 

The structure of $\alpha$-quartz is a network of ordered corner-sharing tetrahedra\cite{Zachariasen1932,Hobbs1996,Marians1990,Buchner2014}. It is believed that the piezoelectric properties are enhanced by the distortion of the structure, which can be represented by two interrelated angles: the bridging angle, $\theta$, between the neighboring tetrahedra and the tilt angle, $\delta$, which is the order parameter of the $\alpha$-$\beta$ quartz phase transition and corresponds to the tilt of the tetrahedra away from its position in $\beta$-phase, where $\delta$=0  \cite{El-Kelany2014,Philippot1996}. The quartz family includes compounds with formulae $TO_2$ (T=Si, Ge) and $MXO_4$ (M=B,Al, Ga,Fe,Mn and X=P, As). Among them, $GeO_2$ and $GaPO_4$ have the most distorted structure, with $GeO_2$ displaying a $d_{11} =6.2 pC/N$\cite{Papet2019HighCrystal,Labeguerie2010} compared with the $2.31 pC/N$ of $SiO_2$\cite{Bechmann1958}. Theoretical calculations of a quartz-type $GeO_2$ SAW device estimate an electromechanical coupling coefficient $K^2$ that can reach a maximum value of 0.34\%\cite{Taziev2018SAWCrystal}, compared to the 0.16\% of the well-known ST-cut $SiO_2$ crystals\cite{Slobodnik1976SurfaceMaterials}.

In the $Si_xGe_{1-x}O_2$ solid solution, it is expected that its structure, and thus the piezoelecric properties, can be tuned continuously via varying the Si/Ge ratio\cite{El-Kelany2014,Lignie2011,Lignie2012,Armand2011}.
Moreover, $GeO_2$ quartz has better thermal stability than $SiO_2$ quartz. The piezoelectric properties of $SiO_2$ start to deteriorate about 300 $^{\circ}C$\cite{Haines2002b}, due to the presence of the $\alpha-\beta$ phase transition at 573 $^{\circ}C$\cite{TuckerKeenD.A.andDoveT.A.2001}, above which the $d_{11}$ of $SiO_2$ vanishes. On the other hand, the $\alpha$-$\beta$ phase transition point of $GeO_2$ is reported to be as high as 1049 $^{\circ}C$ by Sarver and Hummel\cite{Sarver1959}. However, Lignie \textit{et al.} measured $GeO_2$ single crystals by Differential Scanning Calorimetry, showing that $\alpha$-quartz is the only phase present until melting at 1116 $^{\circ}C$\cite{Lignie2011}. This contradiction may arise from size or surface effects: Lignie \textit{et al.} measured $GeO_2$ single crystals while Sarver and Hummel presumably used powders. In addition, a study has shown that nano-crystals of quartz ($SiO_2$) display a continuous $\alpha-\beta$ phase transition, in contrast to the first-order transition in macroscopic quartz ($SiO_2$)\cite{Rios2001}. In any case, the $\alpha$-quartz phase of $GeO_2$ is more stable than that of $SiO_2$ and a study also shows that in the $Si_{0.76}Ge_{0.24}O_2$ composition, the $\alpha$-$\beta$ transition temperature is increased to about 1026 $^{\circ}C$\cite{Ranieri2009}. Moreover, Brillouin spectroscopy measurements of elastic constants up to 1000 $^{\circ}C$ show preservation of piezoelectricity in $GeO_2$ in the measured range\cite{Lignie2014High-temperatureCrystal}. 

Several efforts have been devoted to the growth of bulk single crystals of $Si_xGe_{1-x}O_2$ by hydrothermal synthesis\cite{Ranieri2011,Balitsky2005} and flux growth methods\cite{Lignie2011,Lignie2012,Armand2011}. It was found that when $GeO_2$ crystals are grown in aqueous media, hydroxyl groups catalyze it into the more stable rutile phase\cite{Lignie2011}. Experimental studies show that, at 1200 $^{\circ}C$ and 1 GPa, a maximum of about 40 mole\% Ge can be dissolved into quartz in the bulk\cite{Gullikson2015}. Recently, using an evaporative-recirculation method, Balitsky \textit{et al.}\cite{Balitsky2019EpitaxialCrystals} have succeeded to grow 4-12 mm thick quartz-like $GeO_2$ crystals on $SiO_2$ quartz substrates. Also recently, other works report insights into the crystallization mechanism of $GeO_2$ films\cite{Zhou2021CrystallizationCrystals}, as well as the possibility to obtain improved GeO$_2$/SiO$_2$ miscibility by means of Atomic Layer Deposition multilayers\cite{Antoja-Lleonart2020AtomicMultilayers}. To the best of our knowledge, crystalline thin films of the $Si_xGe_{1-x}O_2$ solid solution have not been reported. 

One of the challenges lies in the solid-state crystallization of the Si-rich part of $Si_xGe_{1-x}O_2$. $GeO_2$ can be readily crystallized into quartz by thermal annealing on various substrates, as recently reported\cite{Balitsky2019EpitaxialCrystals,Zhou2021CrystallizationCrystals}. On the other hand, crystallizing pure $SiO_2$ into quartz is not possible by thermal annealing. It is believed that, besides the small energy differences between the amorphous and the crystalline states, the rigid network of $SiO_2$ tetrahedra introduces a configurational barrier for the crystallization to take place\cite{Hobbs1996,Marians1990}. Interestingly, studies have shown that if the top layers of $SiO_2$ quartz single crystals are amorphized by irradiation with alkali ions, such as $Na^+$, $Rb^+$, $Cs^+$, using ion beam implantation, the amorphous layers can be recrystallized epitaxially by thermal annealing. However, if the irradiating ions are $Si^+$ or $Xe^+$, no recrystallization is observed up to about 900 $^{\circ}C$ \cite{Roccaforte1998,Roccaforte1999,Gustafsson2000,Gasiorek2005,Dhar1999}. Similarly,  Carretero-Genevrier \textit{et al.}, have successfully grown epitaxial thin film of $SiO_2$ quartz on silicon substrates by using $Sr^{2+}$ as catalyst \cite{Carretero-Genevrier2013b,Zhang2019TailoringFilms,Zhang2020Micro/NanostructureSilicon}. In the glass, the alkali and alkali earth ions act as so-called \textit{network modifiers} while $Si^{4+}$ and $Ge^{4+}$ are typical \textit{glass formers}\cite{Bourhis2014}. When theses modifiers are added into the silica network, they break the network by bonding with the non-bridging oxygen. This enhances the mobility of the atoms in the structure and lowers the melting point, as well as the crystallization temperature. 

In this paper, we use pulsed laser deposition (PLD) to first deposit amorphous thin films of $Si_xGe_{1-x}O_2$, which are then crystallized by post-annealing. About 1 at\% Cs is added in some cases to investigate its effect as network modifier. Our study reveals the complexity of the solid-state crystallization of $Si_xGe_{1-x}O_2$ into quartz. We are able to show a larger range of miscibility of $Si_xGe_{1-x}O_2$ solid solution than previously reported with the successful crystallization of (semi-) epitaxial $Si_xGe_{1-x}O_2$ thin films with x ranging from 0 to 0.75 on quartz substrates, which may be interesting for integration in SAW devices.

\section{\label{sec:level1}Experimental}
The process starts with the synthesis of poly-crystalline pellets to be used as targets for laser ablation during PLD growth: Fine powders of $GeO_2$ (Alfa Aesar, 99.9999 \%, $\alpha$-quartz phase) with $SiO$ (Alfa Aesar, 99.8 \%, amorphous) or $SiO_2$ (Alfa Aesar, 99.9 \%, $\alpha$-quartz phase) were mixed together with different ratios. For some of them, a small amount of $Cs_2SiO_3$ was added in. The powders were milled for 90 minutes at 150 rpm and then pressed into a pellet with 20 mm in diameter by a cold pressing at 10 tons. Finally, they were annealed at 900 $^{\circ}C$ for 4 hours. The compositions of the synthesised $Si_xGe_{1-x}O_2$ targets are shown in Table.~S1 of the Supplementary Material. 

Thin films were deposited by pulsed laser deposition using a 248 nm KrF laser (Lambda Physik COMPex Pro 205). The substrates were heated by an infrared laser (DILAS Compact- Evolution, wavelength 808 nm) and the temperature was monitored by a pyrometer. The substrates used in this study are Z-cut (0001) and Y-cut $(10\Bar{1}0)$ $\alpha$-quartz($SiO_2$), $Al_2O_3$ (0001) and $SrTiO_3$ (STO) (001) (CrysTec GmbH). All the substrates are 5 mm x 5 mm in lateral dimensions and 0.5 mm thick. For the epitaxial growth of $GeO_2$, the deposition parameters were: growth temperature of 800 $^{\circ}C$, laser repetition rate of 1 Hz, a total number of 1800 pulses, and cooling rate of 5 $^{\circ}C$ per minute. For the $Si_xGe_{1-x}O_2$ films, first the amorphous thin film was deposited at 600 $^{\circ}C$, with repetition rate of 5 Hz. If not specified, the total number of pulses is 3600. After deposition, the films were annealed at 900 $^{\circ}C$ for 30 minutes and then cooled down with rate of 5 $^{\circ}C$ per minute. The rest of the deposition parameters were the same in both recipes: a target-substrate distance of 46 mm, a laser fluence between 2.5 -3.5 $J/cm^2$. Oxygen was present in the chamber during growth with pressure 0.1 mbar. The annealing atmosphere was 200 mbar of oxygen. The growth process was monitored by high-energy electron diffraction (RHEED) set-up, including differential pumping of the incident electron beam path.

The film crystallinity and orientation, as well as the epitaxial relationship between the thin films and the substrates were studied by x-ray diffraction (XRD) 2$\theta-\omega$ scans and reciprocal space mapping (RSM). The thickness is estimated from x-ray reflectivity (XRR) scans. These studies were performed on a Panalytical X’Pert diffractometer with $CuK_\alpha$ radiation. A hot stage (DHS 900) is used for collecting high temperature XRD data. The morphologies of the thin films were imaged by Atomic Force Microscopy (AFM) (with a Bruker Dimension Icon microscope) and the morphologies of the ablated area of the targets were studied by Scanning Electron Microscopy (SEM) (with a FEI Nova NanoSEM 650 microscope). The stoichiometry of the thin films and the targets before and after laser ablation were studied by Energy Dispersive Spectroscopy (EDS) (with Octane SDD detector by EDAX) and the results were analyzed using the TEAM software. To avoid the charging effects of the insulating targets and thin films during EDS analysis, they were coated with a thin layer (about 20 nm for the targets and 5 nm for the thin films) of Pd/Pt or Au. Local crystallinity and orientation were studied by Electron Backscatter Diffraction (EBSD, with EDAX EBSD system equipped with a Hikari CCD camera). EDAX OIM Analysis 8.1 and Matlab\textsuperscript{\textregistered} based toolbox MTEX\cite{Hielscher2008AAlgorithm} were used for EBSD data analysis. The charging effects during the EBSD data collectrion were suppressed by using low vacuum SEM mode with a pressure of 50 Pa of water vapor.

\section{\label{sec:level1}Results and Discussion}

\subsection{\label{sec:level2}Epitaxial growth of $GeO_2$ on quartz ($SiO_2$) substrates}

\begin{figure}
\centering
\includegraphics[width=0.45\textwidth]{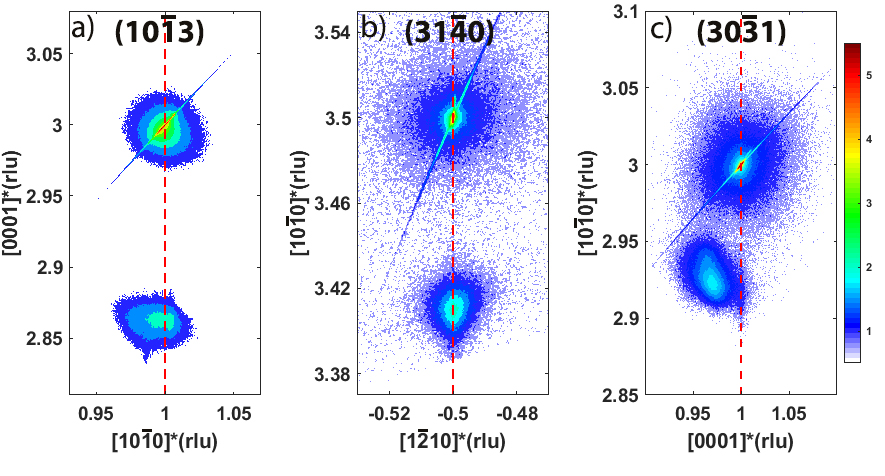}
\caption{\label{fig:epitaxial} Reciprocal Space Maps (RSMs) around (a) the $(10\Bar{1}3)$ Bragg peak, of a Z-cut quartz substrate ([0001] out-of-plane), showing the epitaxially grown $GeO_2$ film; (b) the $(31\Bar{4}0)$ Bragg peak and (c) $(30\Bar{3}1)$ Bragg peak of a Y-cut quartz substrate ($[10\Bar{1}0]$ out-of-plane), showing that the $GeO_2$ film is coherently grown (strained) along the $[1\Bar{2}10]$ direction and relaxed along the $[0001]$ direction. The color bar represents the x-ray intensity (log. scale) in counts per second.}
\end{figure}

We first discuss the growth of the end members of the $Si_xGe_{1-x}O_2$ solid solution. For x=0, $GeO_2$ thin films can grow directly in the quartz phase on quartz ($SiO_2$) substrates at high temperatures. Fig.~S1 (see supplementary material) shows the evolution of RHEED patterns during the deposition of $GeO_2$ on Z-cut and Y-cut quartz substrates. After several tens of pulses, the RHEED pattern turns into grids of dots, suggesting island-like rather than layer-by-layer growth. This is in agreement with the surface morphology obtained by AFM (see Fig.~S2 of the supplementary material), showing that the film is composed of grains of about 30 nm in size. The thicknesses of the films are estimated from XRR scans (Fig.~S3 of the supplementary material) to be 46 nm and 47 nm for Z-cut and Y-cut substrates, respectively. XRD measurements show the thin film has crystallized into the quartz phase and has the same orientation as the substrate in both cases. Indeed, Fig.~\ref{fig:epitaxial}(a) includes an RSM around the $(10\Bar{1}3)$ Bragg peak of Z-cut quartz, which shows that the thin film shares the in-plane lattice parameters of the substrate, therefore, growing epitaxially under strain. The lattice mismatch between the film and the substrate can be estimated as $(a_{SiO_2}-a_{GeO_2})/a_{GeO_2}= (4.916
-4.989)$\AA$/4.989$\AA$=-1.5\% $\cite{Lignie2012}, indicating that the films are under compressive strain. However, for the thin films grown on Y-cut quartz, as shown in Fig.~\ref{fig:epitaxial}(b-c), only the in-plane  $[1\Bar{2}10]$ direction of the thin film grows coherently strained, being the $[0001]$ in-plane direction only partially strained. This is due to the anisotropic lattice mismatch: it is about $-1.5\%$ for the $[1\Bar{2}10]$ direction, while it amounts to about $-4.3\%$ for the $[0001]$ direction\cite{Lignie2012}.

\begin{figure}
\centering
\includegraphics[width=0.45\textwidth]{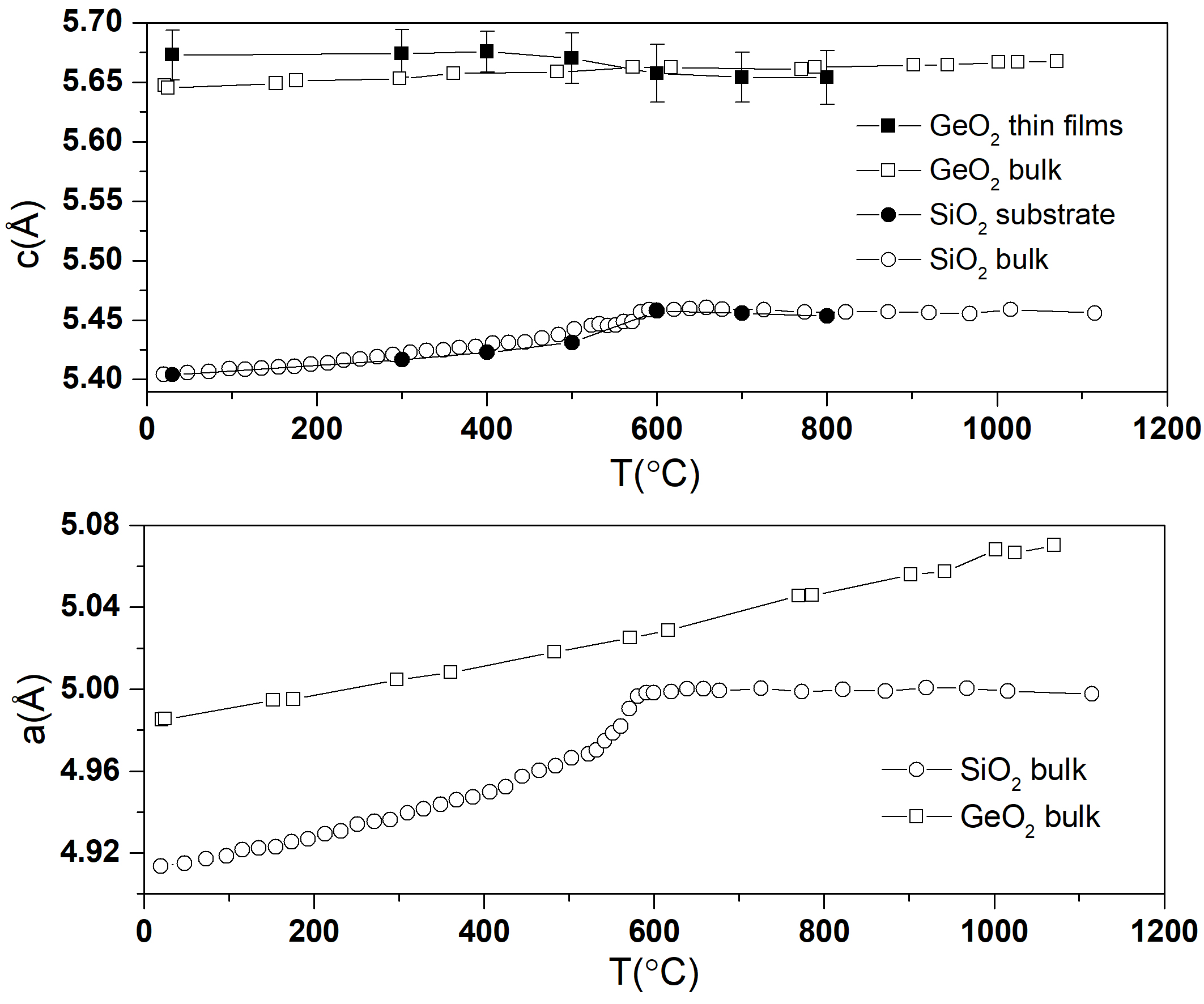}
\caption{\label{fig:temperature} Evolution of lattice parameters of epitaxial thin film of $GeO_2$ on Z-cut quartz substrate (with c-axis, [0001] out-of-plane and a-axis, $[1\Bar{2}10]$ in-plane). The sudden drop of out-of-plane $GeO_2$ film lattice and nearly zero thermal expansion above the $\alpha$-$\beta$ phase transition of the substrate, suggest that the film itself has also undergone an  $\alpha$-$\beta$ phase transition. The error bars of the substrate lattice parameters are smaller than the symbols. The lines are guides to the eyes. The data of bulk $GeO_2$ and $SiO_2$ are obtained from Ref.~39 and Ref.~37, respectively.}
\end{figure}

Fig.~\ref{fig:temperature} shows the change in the lattice parameters of an epitaxial $GeO_2$ thin film grown on a Z-cut quartz substrate, extracted from the $2\theta-\omega$ scans of the $(0003)$ Bragg peak and shown in Table.~S2 (see supplementary material). It can be observed that the $SiO_2$ substrate follows the same temperature dependence as reported in the literature\cite{Carpenter1998CalibrationQuartz}: the lattice of $\alpha$-quartz expands continuously till 573 $^{\circ}C$ where the $\alpha$-$\beta$ phase transition occurs, and the $\beta$-phase shows almost zero and slightly negative thermal expansion\cite{Welche1998NegativeBeta-quartz}. For the $GeO_2$ thin film, the out-of-plane lattice parameter (c, $[0001]$) shows a slight increase for temperatures below about 400 $^{\circ}C$, above which it starts to flatten. However, before the $\alpha$-$\beta$ phase transition of the $SiO_2$ substrates, the out-of-plane lattice of $GeO_2$ thin film is larger than the bulk value. This can be explained by the competition between the thermal expansion and the relieve of in-plane compression from the substrate lattice: as shown in Fig.~\ref{fig:temperature}, bulk crystals of $GeO_2$ show a monotonically increasing lattice with increasing temperature. On the other hand, the in-plane lattice mismatch (a, $[1\Bar{2}10]$) becomes smaller with increasing temperature, leading to a decrease of the out-of-plane lattice expansion. The combination of these two factors results in the complex temperature dependence observed in the $GeO_2$ thin films.

As observed in Fig.~\ref{fig:temperature}, measurements up to 1070 $^{\circ}C$ in bulk $GeO_2$ display a continuous increase of the lattice with temperature, evidencing the absence of an $\alpha$-$\beta$ phase transition\cite{Haines2002a}. This is, however, not the case for the $GeO_2$ thin films. The $GeO_2$ thin film does show signatures of the $\alpha$-$\beta$ transition, following the trend of the $SiO_2$ substrate. We believe this can only be explained by the fact that the $GeO_2$ has transformed into $\beta$-quartz under the influence of the substrate. Under the assumption that the $\beta$-quartz phase of $GeO_2$ has similar zero thermal expansion as $\beta$-quartz $SiO_2$, the volume expansion due to the in-plane lattice mismatch would remain similar. Since the contribution from thermal expansion and volume expansion are both negligible, the lattice of $GeO_2$ thin film after the phase transition of the substrate also remains unchanged. Experimental studies have shown that in the quartz family, only the least distorted homotypes, such as $SiO_2$, $AlPO_4$ and $FePO_4$ undergo $\alpha-\beta$ phase transition\cite{Haines2003ATransition}. It is generally accepted that the $\alpha-\beta$ phase transition is related to phonon mode softening and for the highly distorted homotypes, such as $GeO_2$, the frequency of this mode is temperature independent\cite{Gillet1990High-temperatureHigh-temperatures}. Our findings suggest that, due to the strain imposed by the substrates, the distortion of the $GeO_2$ thin film structure is less than in its bulk form, with the tilt angle $\delta$ close to the $SiO_2$ value. 
 
\subsection{\label{sec:level2}Crystallization of $Si_xGe_{1-x}O_2$}
\subsubsection{\label{sec:level3}Stoichiometric control of the thin films}
Although PLD is known for its good stoichiometry transfer of material from a single target\cite{Schou2009,Ashfold2004}, compared to other thin film deposition techniques, in our study we have found preferential ablation in our $Si_xGe_{1-x}O_2$ targets at all accessible fluence values. As shown in Table. S1 (see supplementary material), the stoichiometry of the $Si_xGe_{1-x}O_2$ targets before and after the laser ablation was studied by EDS. It is clear that for the targets which are made from mixed $SiO_2$ and $GeO_2$ powders, the corresponding thin films lost significant amount of Si comparing to the original composition. Fig.~S4(a) (see supplementary material) shows an example of the ablated area of one of these targets. The ablated surface is not flat but displays columnar features. The images are taken under the Circular Backscatter Detector (CBS), by which the observed contrast mainly comes from the difference in the atomic number of the elements. Thus, the dark contrast observed in the top of the columns indicates that the top is Si-rich. Further EDS analysis (Fig.~S5 in supplementary material) confirms this result, showing that a layer of $SiO_x$ or Si is formed on the top part of the columns.

We believe that the preferential ablation of the targets is caused by the poor laser absorption of $SiO_2$ quartz. Powder XRD (see Fig.~S6 of the supplementary material) has shown that these targets have not formed a solid solution but, instead, they consist of separate crystals of $SiO_2$ and $GeO_2$, both in quartz phase, as expected. Indeed, the $SiO_2$ quartz single crystal is almost transparent to the KrF (248 nm wavelength) laser while $GeO_2$ is somewhat more absorbing: for a $GeO_2$ Z-cut plate with a thickness of 330 $\mu$m, optical transmittance for the KrF laser is about 70 $\%$\cite{Lignie2013}. This leads to the columnar features, as the top layers of less absorbing $SiO_2$ and Si prevent the material under it from being ablated. Due to preferential ablation, we expect our films will have more Ge than the initial composition of the target. The stoichiometry of the films is calibrated by EDS from films deposited on $Al_2O_3$ substrates (to prevent the Si signal arising from the $SiO_2$ quartz substrate) as shown in Table.~S1 (see the supplementary material). In addition, loosely attached particulates are also found on the ablated area due to uneven laser absorption (see Fig.~S7(a) of supplementary material). These particulates can travel to the film surface due to direct ejection from the target\cite{Arnold1999StoichiometryTargets}. With increasing silicon content, the density of the particulates on the thin film surface increases, while $GeO_2$ films are free from particulates. 

To solve these problems, the $SiO_2$ powder was replaced by $SiO$ powder in one of the mixture targets and the improvement is clear: the stoichiometry of the thin films grown from this new target is comparable to that of the target (See Table.~S1 in supplementary material). This is due to the fact that, after synthesis, the amorphous $SiO$ powder has crystallized into the cristobalite phase (Fig.~S8 in supplementary material), which displays stronger absorption for the 248nm laser wavelength. Fig.~S4(b) (see supplementary material) shows the morphology of this target after laser ablation under the CBS SEM detector, where the dark flat pockets are $SiO_2$-rich and the brighter surroundings are the mixture of $SiO_2$ and $GeO_2$. Obviously, the overall morphology is much flatter, with signs of melting and re-solidification, compared to the targets made from $SiO_2$ powder. Moreover, the particulates on the thin films are significantly reduced (Fig.~S7 in supplementary material) due to the better laser absorption. 

\subsubsection{\label{sec:level3}$Si_xGe_{1-x}O_2$ on quartz ($SiO_2$) substrate}

A series of $Si_xGe_{1-x}O_2$ thin films with compositions x= 0, 0.08, 0.16, 0.75, 1 (for Cs-free targets) and x= 0.20, 0.21 and 0.48 (for Cs-doped targets) are deposited at 600 $^{\circ}C$ on Y-cut quartz($SiO_2$) substrates. Surprisingly, no difference between thin films doped with Cs or without it on quartz substrates has been observed. Therefore, in this section we will discuss them together. 

After the deposition, RHEED (see Fig.~S9 of supplementary material) shows no patterns, suggesting that the films are amorphous, as expected at the relatively low temperatures used (see Methods). Then, the thin films are annealed at 900 $^{\circ}C$, where the Cs is reported to become activated as network modifier\cite{Roccaforte1999}. After annealing for 30 minutes, the RHEED pattern is recovered for all compositions, except for the $SiO_2$ thin films (Fig.~S9 in supplementary material), suggesting the amorphous films have crystallized into the $\alpha$-quartz phase with the same orientation as the substrate.

\begin{figure}
\centering
\includegraphics[width=0.45\textwidth]{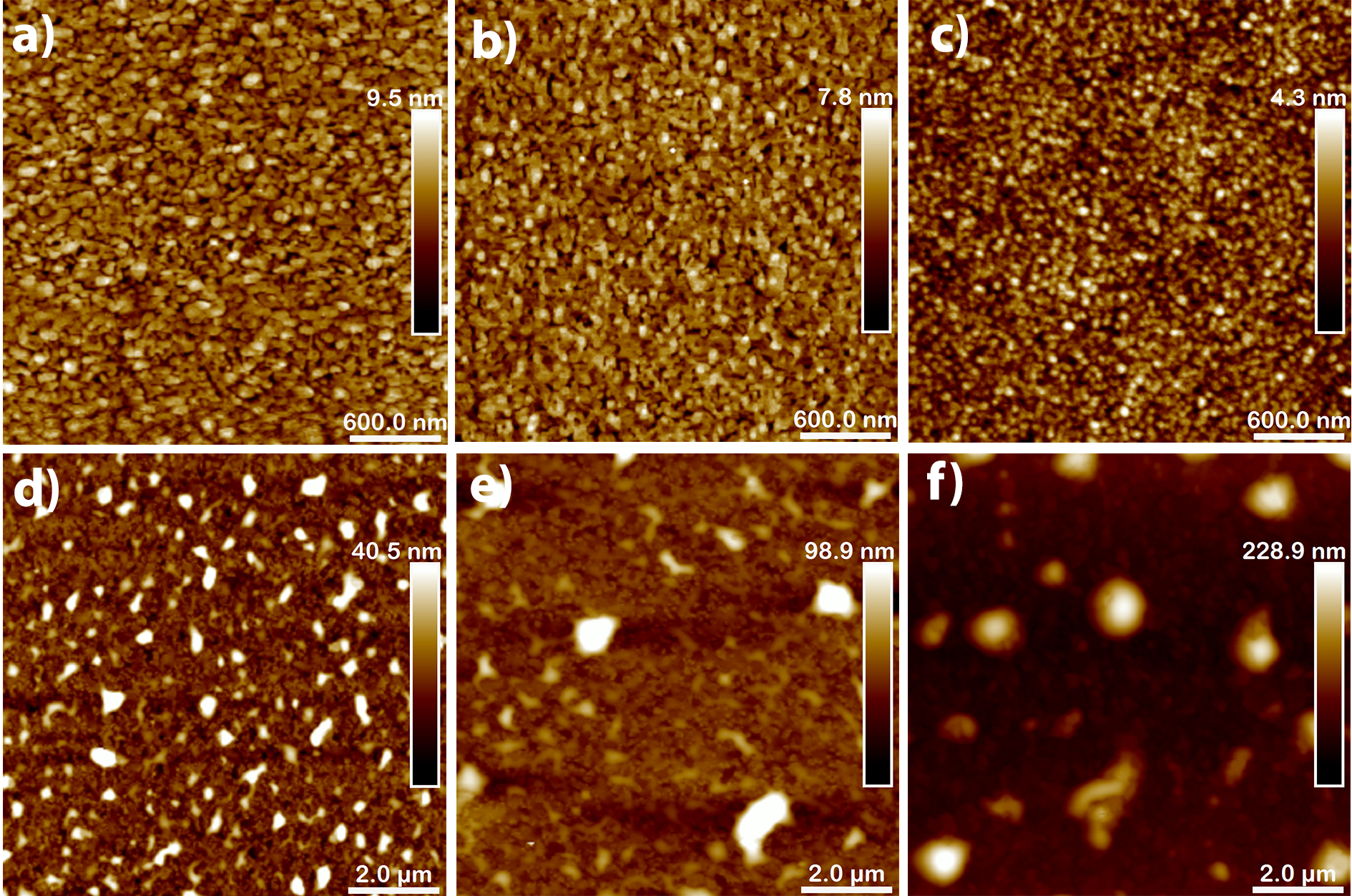}
\caption{\label{fig:AFM} AFM images of $Si_xGe_{1-x}O_2$ thin films thin films grown on Y-cut quartz substrates, with (a) x= 0, (b) x= 0.08, (c) x= 0.16, (d) x= 0.20, (e) x= 0.21 and (f) x= 0.48. Note the columnar features growing in number and height with increasing Si content.}
\end{figure}

For the thin films of $Si_xGe_{1-x}O_2$ (0 $\leq$x$\leq$ 0.48), the crystallization takes places in a similar manner. Fig.~\ref{fig:AFM} shows the morphology of these films: For the $GeO_2$, $Si_{0.08}Ge_{0.92}O_2$ and $Si_{0.16}Ge_{0.84}O_2$ compositions, the films are composed by small grains of about 50 to 150 nm in diameter and the surfaces are flat with root mean square surface roughness ranging from 0.9 to 1.5 nm. With increasing Si amount, such as in $Si_{0.20}Ge_{0.80}O_2$ and $Si_{0.21}Ge_{0.78}O_2$, columnar features with height varying from 50 to 100 nm are present, while the rest of the surface displays small grains, similar to the previous two compositions (see Fig.~S10 of supplementary material). With further increase of Si content, as in $Si_{0.48}Ge_{0.52}O_2$, the height of the columnar features increases to about 150 nm, as shown in Fig.~\ref{fig:AFM}(f). For all these films, the crystallization is complete and the described morphology is homogeneous across the substrate. 

\begin{figure}
\centering
\includegraphics[width=0.45\textwidth]{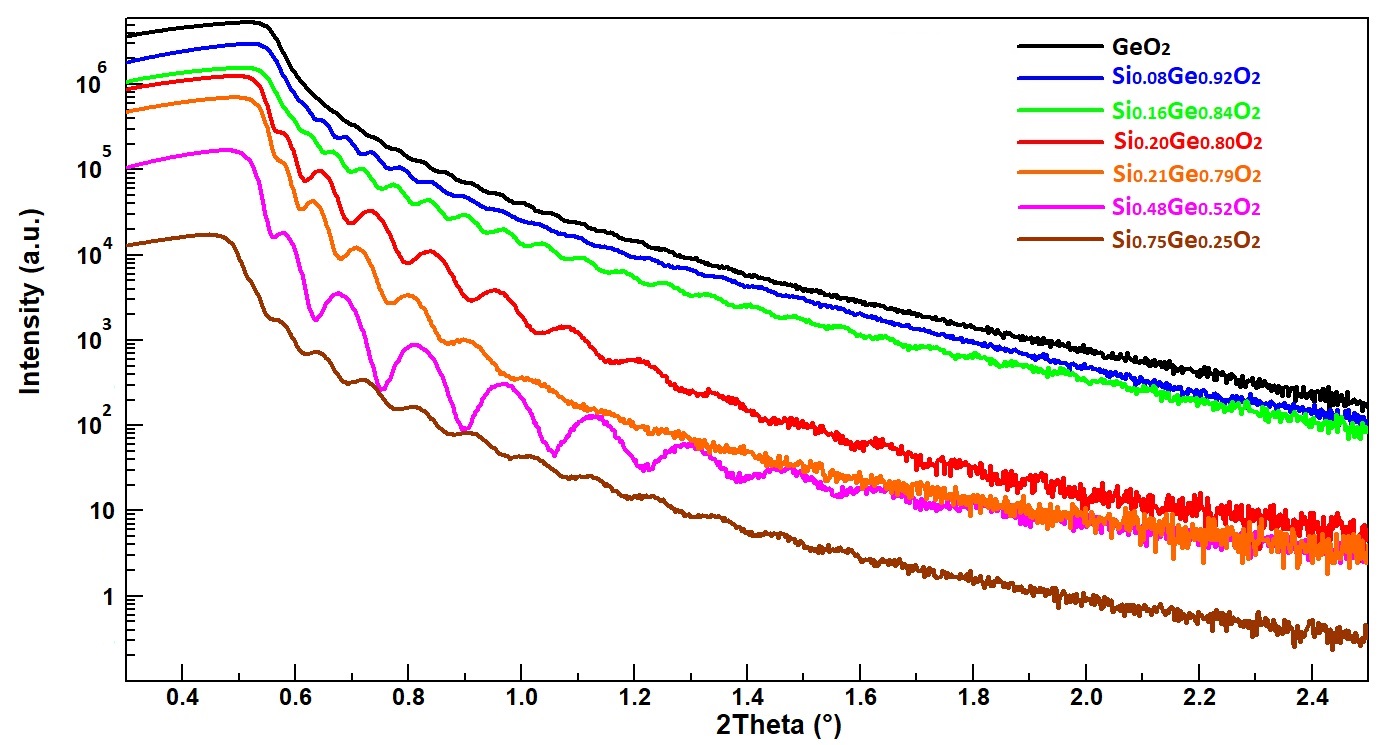}
\caption{\label{fig:reflectivity} X-ray reflectivity (XRR) curves of $Si_xGe_{1-x}O_2$ thin films grown on Y-cut quartz substrates, from which the film thickness and electron density are extracted. They show that the thicknesses of the thin films decrease with increasing silicon content, as a result of the displacement of the material on to the columns. The electron density change of the series can also be easily observed from the shift of the critical angles. The thickness is estimated to be 151 nm, 120 nm, 106 nm, 65 nm, 74 nm, 47 nm and 77nm, for x=0, 0.08, 0.16, 0.20, 0.21, 0.48 and 0.75, respectively.}
\end{figure}

This evolution in morphology is also reflected in the XRR curves as shown in Fig.~\ref{fig:reflectivity}. The film thickness is calculated from period of the Kiessig fringes. While all the films are deposited with the same total number of laser pulses, those with $Si_{0.20}Ge_{0.81}O_2$, $Si_{0.21}Ge_{0.78}O_2$ and $Si_{0.48}Ge_{0.52}O_2$ compositions, are significantly thinner than the rest. This can be attributed to the fact that part of the deposited material is used to form the columns shown in the Fig.~\ref{fig:AFM}. The presence of these taller features also increases the roughness of the film resulting in a quicker drop in intensity at higher angles of the reflectivity curve. The critical angles, which are determined by the electron density of the thin films, decrease with the increase of silicon content, as expected, which suggest the thin films are a homogeneous mixture of $Si_xGe_{1-x}O_2$.  

\begin{figure}
\centering
\includegraphics[width=0.45\textwidth]{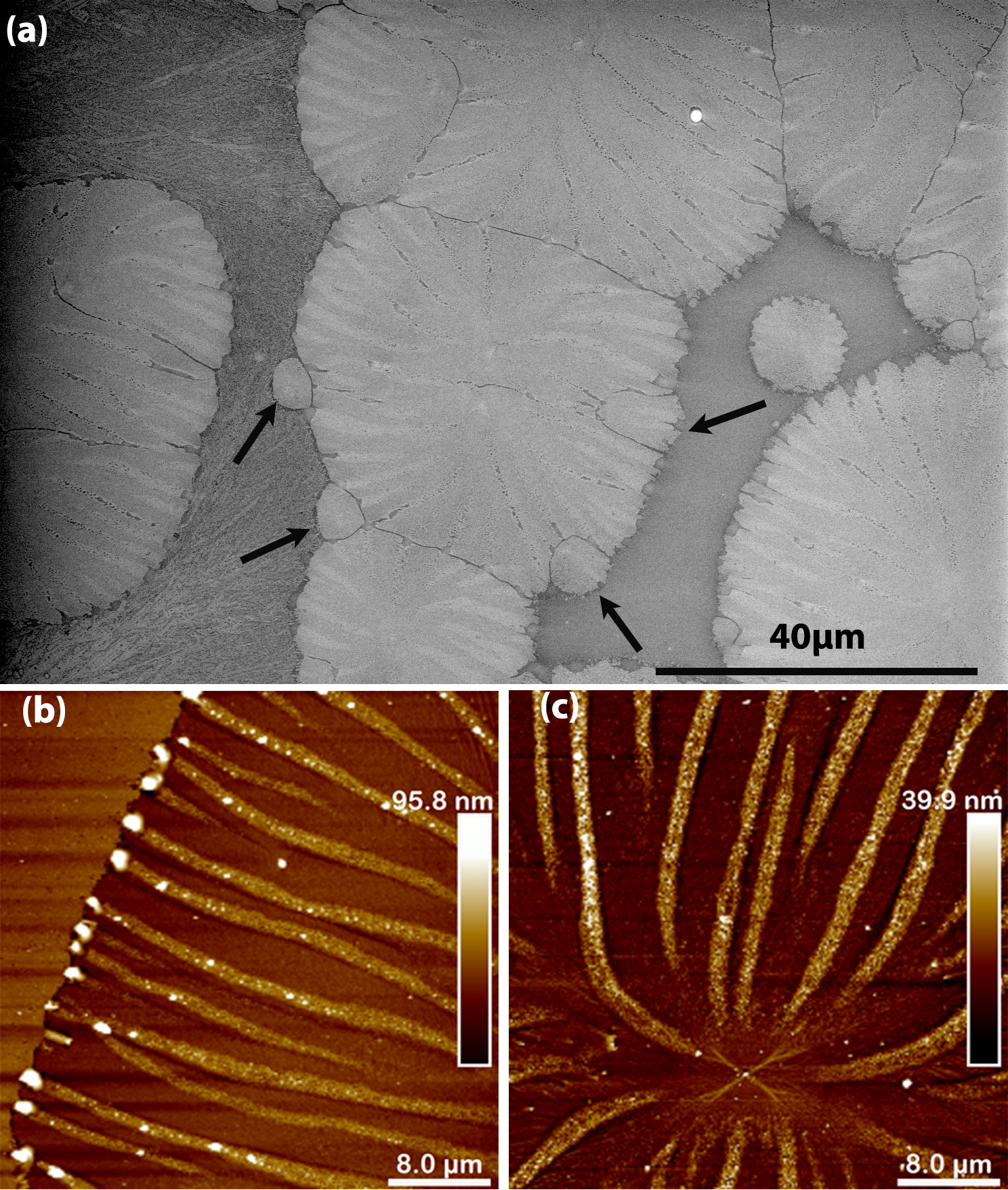}
\caption{\label{fig:domains} (a) SEM images of a $Si_{0.75}Ge_{0.25}O_2$ film grown on Y-cut quartz substrate shows the crystallization of domains surrounded by low crystallinity areas (i) and amorphous areas (ii). The black arrows point to the new domains nucleated at the edge of a bigger domain. (b) AFM image of a domain boundary where bumps are formed when the stripes meet the boundary. Notice the stripes are about 10 nm higher than the areas in between them. (c) AFM image showing the middle part of the domains with the nucleation point is a central point in area from which stripes arise.}
\end{figure}

On the other hand, the $Si_{0.75}Ge_{0.25}O_2$ composition shows a different type of crystallization. As shown in Fig.~\ref{fig:domains}, unlike in the previous films, only about 90\% of the sample area is crystallized. Moreover, instead of giving rise to a continuous crystallized film with small grains, it is crystallized into oriented domains. AFM images in Fig.~\ref{fig:domains}(b) and (c) show that nucleation starts at the center of the domain. After some further crystallization, stripes with width of about 1 $\mu$m form and expand radially. In between these stripes, there are flat and dense crystalline areas, which are about 10 nm lower in height than the stripes. When these stripes reach the domain boundary, bumps as high as 100 nm form at the end of the stripes, suggesting accumulation of excess material. Interestingly, the stripes have the same height as the area outside of the domain in Fig.~\ref{fig:domains}(b). SEM images with higher magnification show that the stripes have a porous structure and are composed of small holes about 50 nm in diameter (See Fig.~S11 in supplementary material). 

\begin{figure}
\centering
\includegraphics[width=0.45\textwidth]{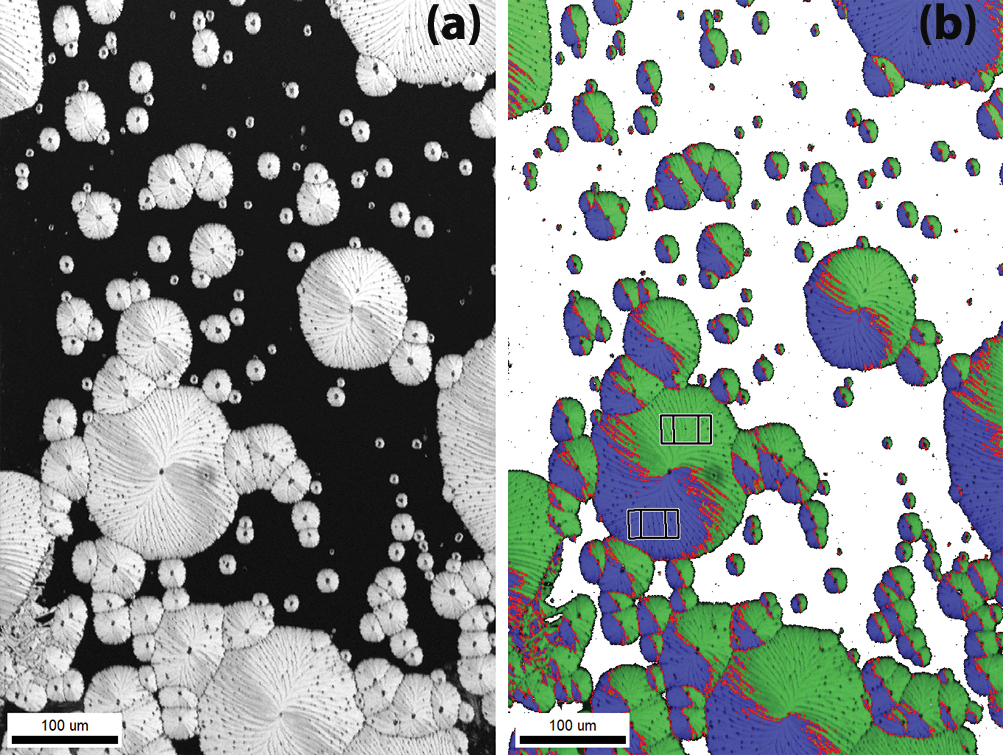}
\caption{\label{fig:twins} EBSD analysis of domains on $Si_{0.75}Ge_{0.25}O_2$ thin films. (a) Image Quality (IQ) map where the grey-scale represents the degree of crystallinity. (b) Quartz crystal orientations map showing two observed orientations: $(\Bar{1}010)[0001]$ (blue) and $(\Bar{1}100)[0001]$ (green), combined with the IQ signal. All quartz domains are composed of Dauphin\'{e} twins and the red lines are the twin boundaries. Notice the shape of the twin boundaries is dependent on the size of the domains: For small domains, the twin boundaries are nearly vertical lines; while with further growth of these domains, the boundaries rotate counterclockwise, finally forming a "Yin-Yang pattern".
}
\end{figure}

These kind of radial structures and circular domains resemble the spherulitic crystallization of $GeO_2$ thin films on $Al_2O_3$ reported in our previous study\cite{Zhou2021CrystallizationCrystals}. However, after further examination by EBSD, it turns out that these domains are single crystals as shown in Fig.~\ref{fig:twins}, which agrees with the oriented RHEED pattern collected for the same films. Furthermore, GIXRD did not show any peaks confirming the epitaxial nature of the domains (see Fig.~S12(a) in supplementary material). Moreover, some of the areas surrounding the domains (marked as i) in Fig.~\ref{fig:domains}(a)), appear not to be amorphous, but instead they display poly-crystals with poor crystallinity, as it is shown by their vague Kikuchi patterns. In fact, the crystallinity of this area is so low that it cannot be indexed properly by EBSD. Comparing to area i) in Fig.~\ref{fig:domains}(a), where substructures with different contrast can be observed (See Fig.~S11 in supplementary material for higher magnification), area ii) has a homogeneous contrast, suggesting that this area is amorphous. 

Fig.~\ref{fig:twins}(a) is the Image Quality (IQ) map of a EBSD scan, where the contrast represents the crystallinity: the brighter the contrast, the higher the crystallinity. The stripes can be clearly observed in the IQ map and they have worse crystallinity than the rest of the area on the domains. Interestingly, from Fig.~\ref{fig:twins}(b), all the domains are composed of Dauphin\'{e} twins, as clearly indicated by the blue-green color in the orientation map. The Dauphin\'{e} twins detected by EBSD are related to each other by a 60 $^{\circ}$ rotation around the c-axis, such as the two unit cells drawn in Fig.~\ref{fig:twins}(b). By comparing the pattern in the domains in Fig. \ref{fig:twins}(a) and (b), it can be observed that the twin boundaries are related to the stripes as sometimes the twin boundaries overlap with the stripes. SEM images show these stripes are well-organized structures as they can connect to each other across the domain boundaries (See Fig.~S11 in Supplementary material). Moreover, it can be observed that the twin boundaries evolve with the growth of the domains: For the smallest domains, the twin boundaries are straight lines close to the 11-5 o'clock direction. With further growth of the domain, the twin boundaries are still straight but rotate counter-clockwise, eventually forming a "\textit{Yin Yang} pattern" with a jerky Dauphin\'{e} twin boundary in the big domains (See Fig.~S13 in Supplementary material for another example with higher magnification).

Dauphin\'{e} twins can form during the $\beta$-quartz to $\alpha$-quartz phase transition, as growth twins, or when the crystal is under mechanical stress, as mechanical twins\cite{Menegon2011TheQuartz,Wenk2006DauphineNovaculite,Minor2018TrackingQuartzite}. $\alpha$-Quartz crystal is highly anisotropic in stiffness, with the stiffest orientation being close to negative rhombohedral (z) $(01\Bar{1}1)$ with a Young's modulus of 109 GPa, while the positive rhombohedral (r) $(10\Bar{1}1)$ has a Young's modulus of about 71.5 GPa\cite{Minor2018TrackingQuartzite}. It is generally accepted that when a constant stress is applied on a quartz domain, the twins will react in a way to maximize the stored elastic strain energy of the domain in a way to minimize the internal energy of the system, and the stored elastic energy is proportional to the difference between the elastic energies of the two parts of the twins\cite{Thomas1951PiezoerescencetheStress,Tullis1972PreferredExperiments}. Compression studies have shown that if a compressive stress is applied perpendicular to the rhombohedral (z) planes, twins will form. However, if it is applied perpendicular to the rhombohedral (r) planes, twinning will not form, since it is already in a thermodynamic stable state \cite{Minor2018TrackingQuartzite,Menegon2011TheQuartz,Tullis1972PreferredExperiments}. 
We believe the Dauphin\'{e} twins in our films are induced by the epitaxial stress from the substrates because of their highly ordered structures. However, in our case, the crystal orientation is fixed by the epitaxial growth and only twin boundaries can move. As domains grow bigger, the accumulated stress due to the epitaxial relationship increases with increasing domain size. Moreover, the accumulated stress is not equal in the two in-plane directions, with the compressive strain along [0001] being about 4 times larger than that along the $[1\Bar{2}10]$ direction, as will be explained in the later sections.
We believe this is the reason for the changing of twinning patterns with domain growth. The details of the mechanism will be discussed in a separate paper.

For the $SiO_2$ thin films made using the same annealing procedure, the RHEED pattern is not recovered, suggesting they are amorphous. Unfortunately, it is difficult to check this by XRD because if crystallized, the film peak will overlap with the substrate peak. Under the optical microscope, the film looks homogeneous. without any features, and the AFM scans show the surface is flat with $R_q$ = 0.3 nm. Moreover, an XRR scan shows an uniform layer with density of about 2.2 $g/cm^3$ is grown on the substrate, to be compared with the 2.64 $g/cm^3$ value of $SiO_2$ quartz (see Fig.~S14 in the supplementary material). All this evidence supports the amorphous nature of the $SiO_2$ thin films.

\begin{figure*}
\centering
\includegraphics[width=1\textwidth]{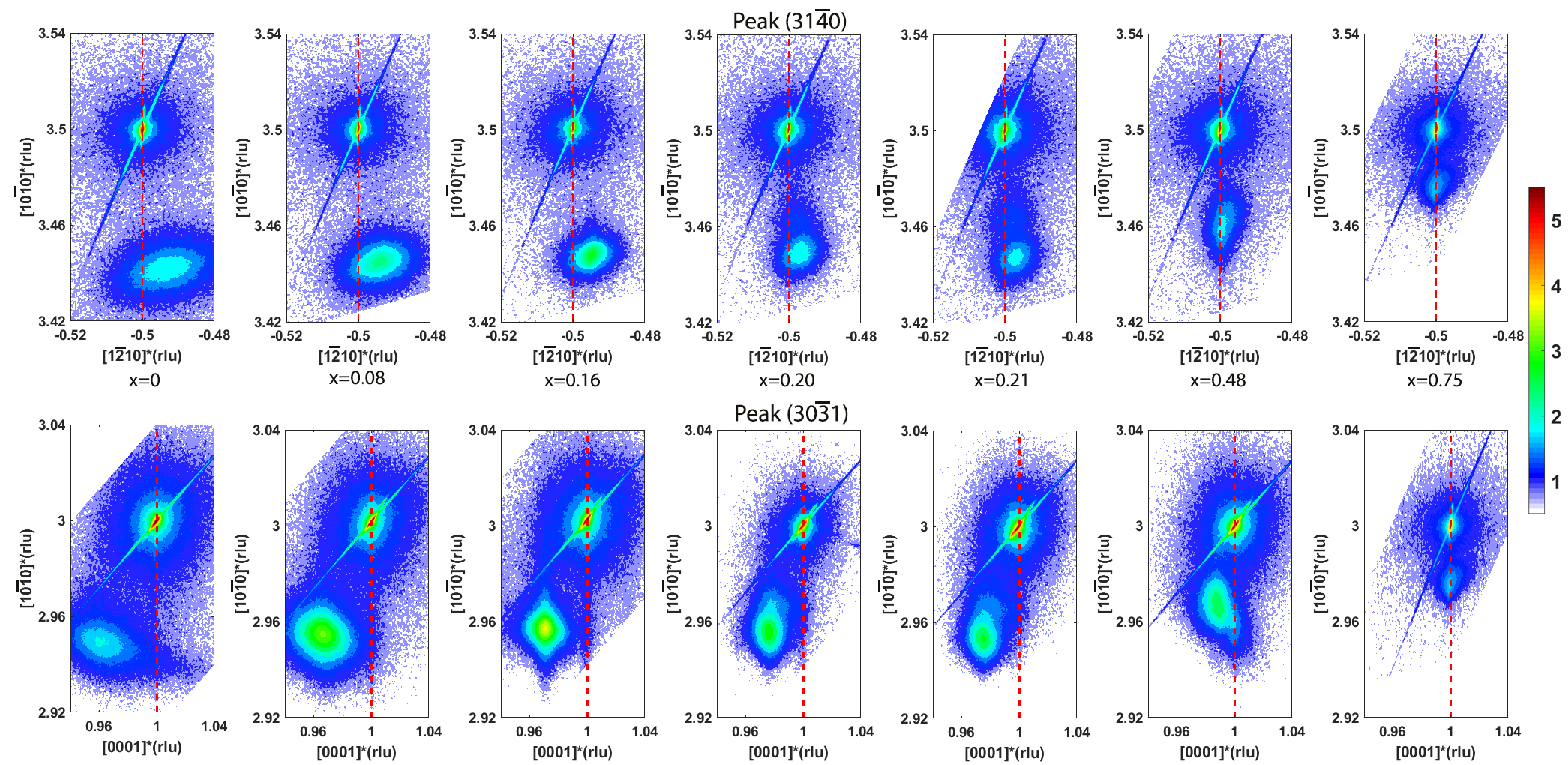}
\caption{\label{fig:spacemap} Reciprocal Space Maps (RSM) around the $(31\Bar{4}0)$ peak (top row) and $(30\Bar{3}1)$ peak (bottom row) of the Y-cut quartz substrates also displaying the $Si_xGe_{1-x}O_2$ film peaks. The vertical axes represent the Y-direction and the horizontal axes represent the X- and Z- directions in the top and bottom rows, respectively. The color bar indicates the x-ray intensity (log. scale) in counts per second.}
\end{figure*}

\begin{figure}
\centering
\includegraphics[width=0.45\textwidth]{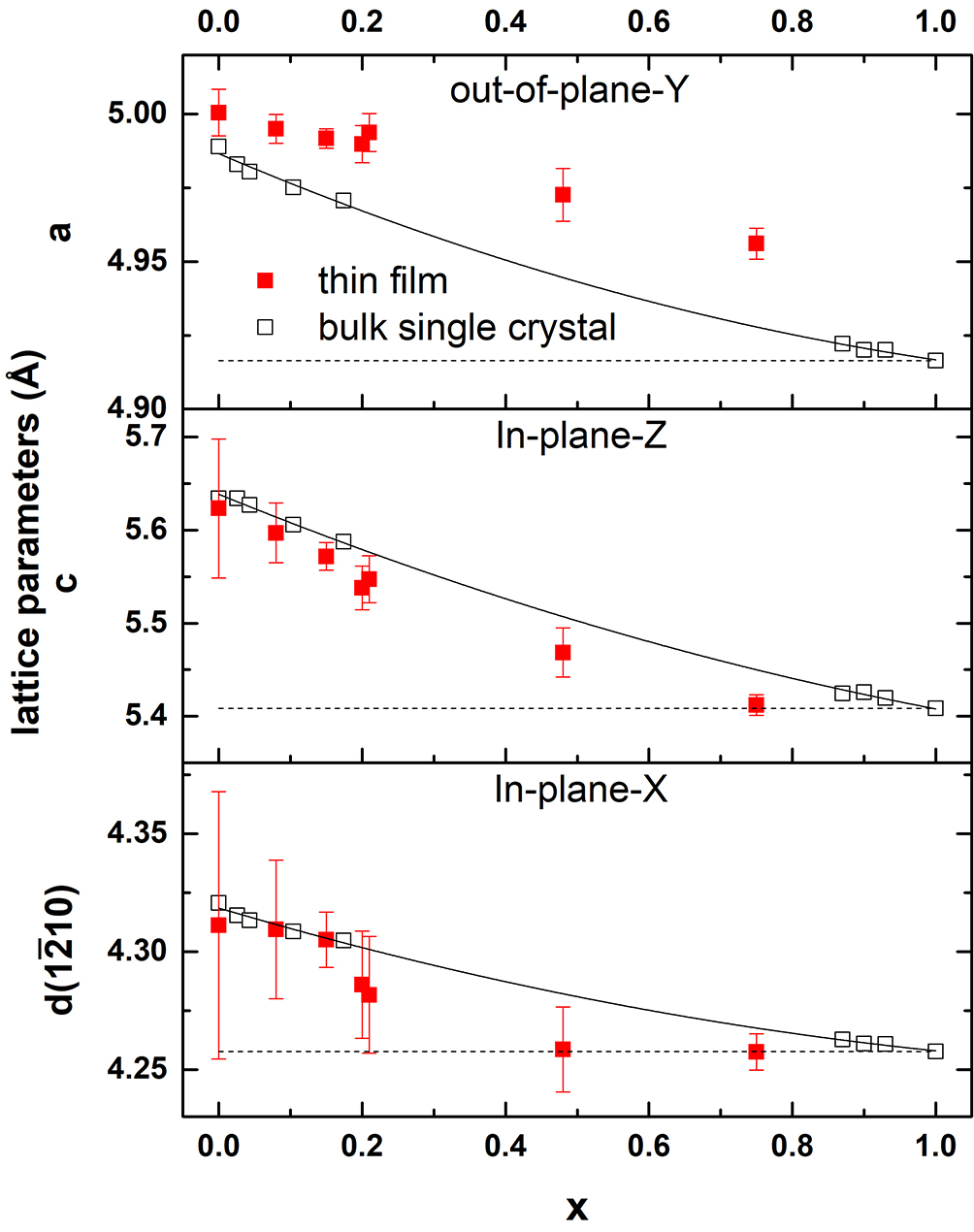}
\caption{\label{fig:lattice} Evolution of lattice parameters of the $Si_xGe_{1-x}O_2$ solid solution grown on Y-cut quartz substrates. The thin film data (solid squares) are obtained from Fig.~\ref{fig:spacemap} (also collected in Table.~S3). The data from bulk single crystal are obtained from Ref.~17 and Ref.~25. The solid lines are the second-order polynomial fitting of the bulk data, as described in the Ref.~17. Dash lines are the lattice parameters of the substrates. The thin film lattice in the [0001] direction has similar composition dependence as the bulk single crystals, while the $[1\Bar{2}10]$ direction is more strongly subjected to the stress from the substrate, being nearly fully strained for x=0.48. The compressive strain in the $[1\Bar{2}10]$ direction is mainly accommodated by the expansion in out-of-plane direction.}
\end{figure}

Fig.~\ref{fig:spacemap} displays X-ray RSMs around the $(31\Bar{4}0)$ and $(30\Bar{3}1)$ Bragg peaks of the Y-cut quartz substrates, in which the peaks of the $Si_xGe_{1-x}O_2$ films are also visible. Table.~S3 (See supplementary material) contains the thin film lattice parameters that are extracted from those maps by means of Lorentzian fitting (see Fig.~S15 of the supplementary material for one example of the procedure), which are also plotted in Fig.~\ref{fig:lattice}. In this figure, the bulk lattice parameters for different compositions, obtained from Ref.~17 and Ref.~25, are also plotted. According to these results, none of these thin films are fully relaxed: the in-plane lattice parameters are smaller than those of the bulk single crystals due to the in-plane compressive stress applied by the substrates. Consequently, the out-of-plane lattice is elongated with respect to the bulk value. 

From Fig.~\ref{fig:spacemap}, it can be observed that, for the thin films with low Si content (x=0, 0.08 and 0.16), the film peaks shift towards the substrate peaks in the out-of-plane direction $[10\Bar{1}0]$ and both in-plane directions,$[1\Bar{2}10]$ and [0001], with increasing Si content, as expected due to the larger unit cell volume of $GeO_2$. However, with x increasing further (x= 0.20 and 0.21), the in-plane lattice keeps decreasing while the out-of-plane lattice spacing remains comparable to that of the film with x=0.16. This can be explained by the very large compressive strain along the in-plane $[1\Bar{2}10]$ direction, as shown in Fig.~\ref{fig:lattice}, where the lattice drops dramatically from x=0.16 to 0.20 in the $[1\Bar{2}10]$ direction. In response of that, the lattice expands on the strain-free, out-of-plane direction.

With further increase of Si content to x= 0.48, the thin films are fully strained by the substrate in the $[1\Bar{2}10]$ direction as shown in Fig.~\ref{fig:spacemap}. Interestingly, in the [0001] direction, the tail of the thin film peaks reaches the substrate indicating that a part of the film is fully strained. Using the lattice parameter data from the bulk single crystals at various compositions\cite{Lignie2012}, we can estimate the strain for the x=0.48 thin film to be about -0.7\% and -2.2\% in $[1\Bar{2}10]$ and [0001] directions,respectively. For the first few layers, the strain imposed by the substrate forces the film to have the same lattice as the substrate both in the $[1\Bar{2}10]$ and [0001] directions. However, with the crystallization continuing upwards, the elastic energy contributed along the [0001] direction is too large to be accommodated, relaxing the lattice gradually towards the bulk values, and contributing to the main intensity of the $(30\Bar{3}1)$ peak. On the other hand, in the $[1\Bar{2}10]$  direction, the mismatch strain of about -0.7\% can be accommodated and the lattice is completely strained for the entire thickness. 

Finally, when x increases to 0.75, the $Si_{0.75}Ge_{0.25}O_2$ thin films are fully strained by the substrate in both in-plane directions, as shown in Fig.~\ref{fig:spacemap}. Using similar arguments, we can estimate the strain magnitudes to be -0.2\% and -0.8\% along the $[1\Bar{2}10]$ and [0001] directions, respectively. From Fig.~\ref{fig:lattice}, it can be observed that, generally speaking, the lattice along the [0001] direction has a composition dependence more similar to the bulk than the lattice in $[1\Bar{2}10]$ direction, due to the strain anisotropy described above.

\subsubsection{\label{sec:level3}$Si_xGe_{1-x}O_2$ on Sapphire ($Al_2O_3$) substrates}

\begin{figure}
\centering
\includegraphics[width=0.45\textwidth]{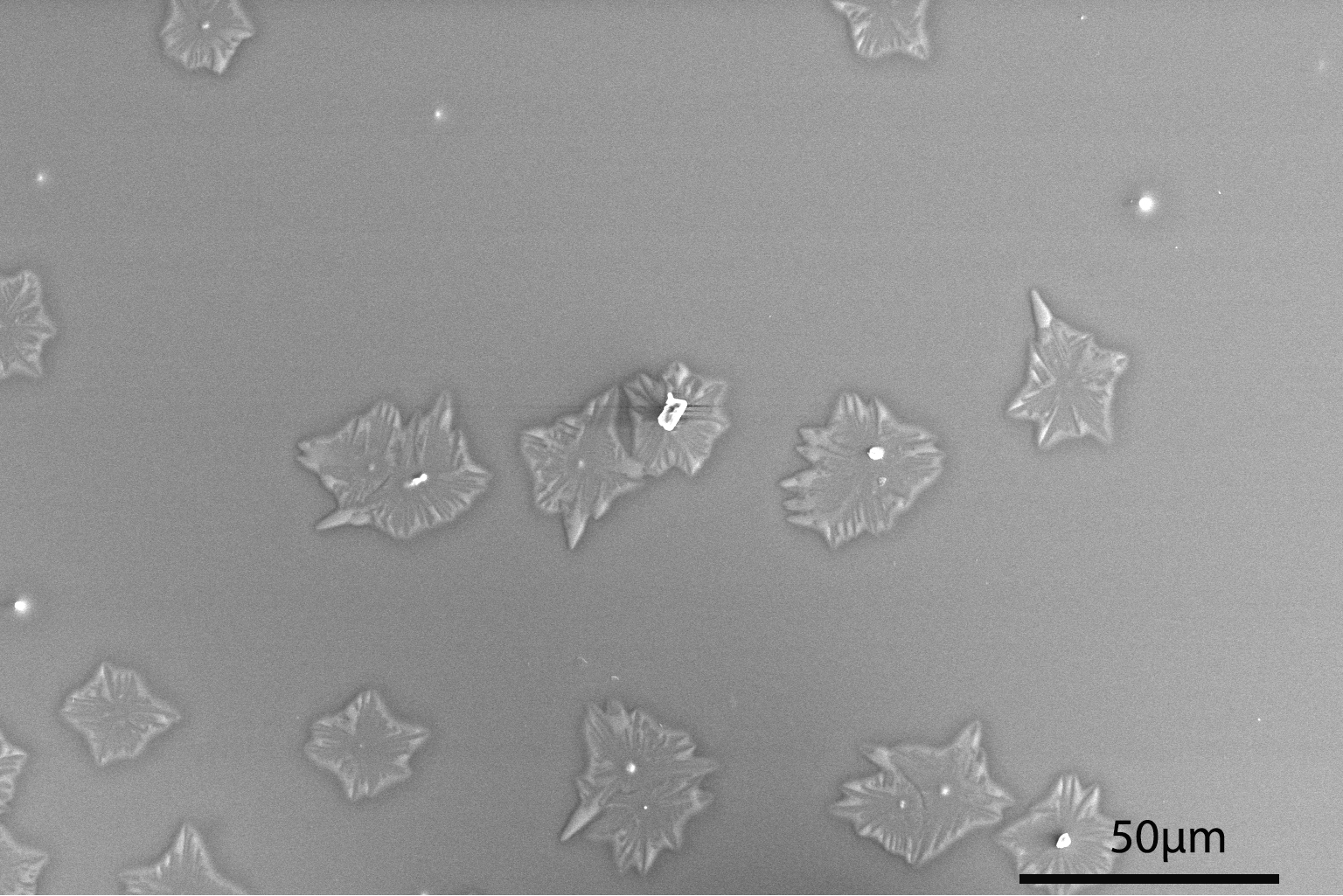}
\caption{\label{fig:SEM} SEM image of a $Si_{0.08}Ge_{0.92}O_2$ film on sapphire substrates showing that some crystals are formed on the thin film}
\end{figure}

A series of $Si_xGe_{1-x}O_2$ films with x= 0.08, 0.16, 0.75, 1 (for Cs-free targets) and x= 0.20, 0.21, 0.48 (for Cs-doped targets) were also deposited on $Al_2O_3$ substrates with the same growth parameters as those used for the growth on quartz substrates (see previous section). In our earlier study\cite{Zhou2021CrystallizationCrystals}, we have reported crystallization of $GeO_2$ on $Al_2O_3$ substrates: quartz crystals with diameter of about 100 $\mu$m are achieved, but most of the crystalline area is covered by wavy patterns that start from the edge of the sample and sweep inwards until crystallization of the entire film is completed. GIXRD shows that these crystals are randomly oriented poly-crystals (See Fig.~S16 in supplementary material). However, these continuous wavy features are absent in the Si-containing $Si_xGe_{1-x}O_2$ films. Fig.~\ref{fig:SEM} shows the $Si_xGe_{1-x}O_2$ thin film with x=0.08 after annealing. Some quartz poly-crystals with diameter of about 20 $\mu$m are present while the rest of the film is still amorphous. Films with x=0.16 show a similar behavior, with the difference of having smaller crystals with diameters below 10 $\mu$m (See Fig.~S17 in supplementary material). The decrease of the size of the quartz crystals with increasing Si content indicates an enhancement of the crystallization barrier, as expected. 

EDS point analysis on the crystals as well as the amorphous area results in the same composition (See Fig.~S17 in supplementary material) suggesting that the crystals are solid solutions of $Si_xGe_{1-x}O_2$. As observed from Fig.~\ref{fig:SEM} and Fig.~S18 (see supplementary material), the center of the crystal, the nucleation point, can be as flat as the surrounding area or it can protrude above the surface. Moreover, in the amorphous area, similar round bumps are observed, perhaps denoting nucleation points for the new crystals that have not grown yet. Interestingly, EDS point analysis on these cores and bumps often shows a different stoichiometry than the nominal x value (Ge/Si ratio) of the thin film, while the rest of the film does show the expected stoichiometry (Fig.~S17 in supplementary material) .  

For the rest of the $Si_xGe_{1-x}O_2$ thin films with x=0.20, 0.21, 0.48 (for Cs-doped targets) and x=0.75, 1 (for Cs-free targets), no crystallization is observed on the $Al_2O_3$ substrates (See Fig.~S13(b) in supplementary material). This may indicate that the crystallization barrier is so high that 900 $^{\circ}C$ is not sufficient or the amount of Cs (about 1 at\%) is not enough, to fully break the network. 

\subsubsection{\label{sec:level3}Crystallization mechanism of $Si_xGe_{1-x}O_2$}

We have shown that the films on quartz and sapphire substrates display different morphologies: the thin films on quartz substrates are composed of small grains, while on sapphire substrates, they crystallize into more extended crystals with clear nucleation points. Moreover, we have found that changing the deposition temperature from 600 $^{\circ}C$ to 800 $^{\circ}C$, $GeO_2$ thin film can still crystallize successfully on the quartz substrates. However, this is not the case on sapphire substrates. On these substrates, the thin film starts evaporating from the edge of the sample, and circular evaporation pits can be observed (See Fig. S19 in supplementary material). AFM images of the film reveal a flat surface without any features, suggesting it is still amorphous (Fig.~S19 in the supplementary material) and GIXRD confirms the amorphous state of the thin film. This evidence indicates that at 800 $^{\circ}C$, the high temperature suppresses the nucleation process of $GeO_2$, leading to no crystallization. 

In the $Si_xGe_{1-x}O_2$ thin films, the crystallization kinetics appears to change with composition. The melting point of $GeO_2$ is 1116 $^{\circ}C$\cite{Lignie2011a}, compared to about 1600 $^{\circ}C$ of $SiO_2$, and the annealing temperature is 900 $^{\circ}C$ in both cases. One would expect that the melting point in the composition series increases with increasing silicon content. As the size of the crystals on sapphire substrates decreases from about 100 $\mu$m (x=0), 20 $\mu$m (x=0.08), to below 10 $\mu$m (x=0.16), this indicates that the growth rate of $Si_xGe_{1-x}O_2$ decreases with increasing Si content in this range, and is likely to decrease further with additional increase of silicon content. On the contrary, the nucleation rate is increasing in this range as the number of crystals on sapphire substrates increases (See Fig.~S20 of supplementary material). When x reaches x= 0.20, the growth rate on sapphire is so low that no crystallization occurs. On the quartz substrates, the growth rate is also likely decreasing with increasing x. Nevertheless, the growth rate is still high enough to fully crystallize the $Si_{0.48}Ge_{0.52}O_2$ thin films. The fact that the $Si_{0.75}Ge_{0.25}O_2$ composition only shows partial crystallization, suggests that at this Si concentration the growth rate is significantly reduced compared to that of $GeO_2$.

Thus, we propose the following crystallizing mechanisms: for the $Si_xGe_{1-x}O_2$ (0 $\leq$x$\leq$ 0.48) thin films on quartz substrates, the substrates can serve as a seed for crystallization. Thus, there is no need for additional nucleation points and the $Si_xGe_{1-x}O_2$ (0 $\leq$x$\leq$ 0.48) can grow on the quartz substrate homo-epitaxially, as also shown for bulk $GeO_2$\cite{Balitsky2019EpitaxialCrystals}. In this case, the whole film crystallizes in a planar and continuous fashion from bottom to top: at the film-substrate interface, the atoms diffuse and attach to the substrate to crystallize, forming small grains, and the strain from the substrate forces these layers to grow epitaxially. When this layer finished crystallization, atoms in the topper layer repeat the process until the film is fully crystallized. 

On the contrary, for the $Si_xGe_{1-x}O_2$ thin films on sapphire substrates, an additional nucleation barrier is present. Our recent study on $GeO_2$ thin films on sapphire substrates has found that the nucleation starts from the top surface of the thin film\cite{Lutjes2021SpheruliticFilms}. Thus, the growth front at the film top is advanced with respect to the growth front at the film bottom, leading to rotation of the lattice during the growth\cite{Lutjes2021SpheruliticFilms}. When the nucleation starts from the top, it loses the information from the substrate and results in randomly oriented crystals. During the deposition at 600 $^{\circ}C$, it is very likely that the nucleation process already starts. In the follow-up annealing at 900 $^{\circ}C$, the increase of growth rate allows the nuclei to grow laterally, forming the crystals shown in Fig.~\ref{fig:SEM}. Besides the nucleation barrier, poor molecular adhesion also makes it difficult for the $Si_xGe_{1-x}O_2$ to crystallize on sapphire. Longer annealing time (e.g. 2 hours) does not help further crystal growth. Instead, it destabilizes the quartz crystal, with material evaporating out, as shown in Fig.~S21 in supplementary material.

Interestingly, $Si_{0.75}Ge_{0.25}O_2$ films on quartz substrates exhibit distinct characteristics: unlike the continuous films on quartz obtained with other compositions, this composition consists of quartz domains with clear nucleation centers. Although this resembles the quartz crystals on sapphire, these domains are single crystals fully strained by the substrates and, thus, clearly distinct from the $Si_xGe_{1-x}O_2$ poly-crystals on sapphire. As mentioned earlier, at x= 0.75, the growth rate is significantly reduced when compared to that at x= 0, and it may not be kinetically favorable for the growth to proceed from the bottom to the top as in the other compositions on quartz. On the other hand, the nucleation rate is still sufficiently high, as suggested by comparing the number of crystals for this composition (Fig.~\ref{fig:twins}) with those in $Si_{0.08}Ge_{0.92}O_2$ films on sapphire (Fig.~\ref{fig:SEM}). Moreover, as pointed by the black arrows in Fig.~\ref{fig:twins}, small new domains nucleated at the edge of the big domains. The epitaxial nature of the film requires for the nuclei to form at the interface between the substrate and the thin film. Once a nucleus is formed, it is more kinetically favorable for the atoms to attach to the existing nuclei and expand laterally forming the circular domains shown in Fig.~\ref{fig:twins}. Due to the small mismatch between $Si_{0.75}Ge_{0.25}O_2$ and the $SiO_2$ substrate, the film is fully strained by the substrate.

\subsubsection{\label{sec:level3}$Si_{0.48}Ge_{0.52}O_2$ and $SiO_2$ on $SrTiO_3$ substrates}

\begin{figure}
\centering
\includegraphics{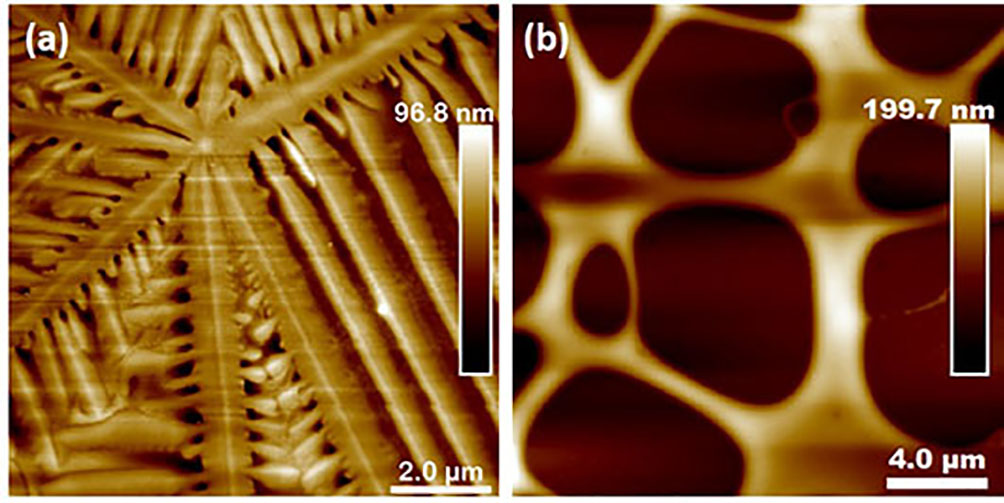}
\caption{\label{fig:STO} AFM images of (a) crystalline $Si{0.48}Ge_{0.53}O_2$ on a $SrTiO_3$ substrate. (b) the evaporation pits observed in the same sample, revealing the bare substrate.}
\end{figure}

Unexpectedly, our experiments on quartz and sapphire substrates shown on the previous sections do not reveal the functionality of Cs in the network. Next to Cs, Sr is reported to act as a network modifier, which can also help crystallization by bonding with oxygen atoms\cite{Antoja-lleonart2021GrowthSubstrates}. In our previous study\cite{Zhou2021CrystallizationCrystals}, we have found that when $GeO_2$ is deposited on $SrTiO_3$ (STO) substrates, during the annealing at 830 $^{\circ}C$, Sr from the substrate will diffuse into the film and react with it forming a colorful germanate $SrGe_{3.3}O_{5.6}$, on top of which a quartz layer forms. To study the role of the network modifiers in the $Si_xGe_{1-x}O_2$ network, we have deposited $Si_{0.48}Ge_{0.52}O_2$ films on STO substrates. After annealing for 30 minutes at 800 $^{\circ}C$, some crystals appear (Fig.~\ref{fig:STO}(a)), accompanied by some circular pits (Fig.~\ref{fig:STO}(b)). Further investigation by EBSD reveals the polycrystallinity of the quartz crystal, and the nature of circular spots are evaporation pits, where all the film material evaporates, revealing the bare substrate surface (See Fig.~S22 in supplementary material). 

Comparing to the lack of crystallization of the same composition on sapphire at 900 $^{\circ}C$, the presence of crystals in this case suggests that, indeed, Sr can break the network and help crystallization. However, if the film thickness is reduced by one-half, the number of the evaporation pits increases significantly, while the number of crystals remains very much unchanged (see Fig.~S23 in supplementary material). Moreover, a gradient for crystallization/evaporation exist across the surface. Close to the edge, the density of crystals and evaporation pits is the highest, with a gradual decrease towards the center of the film.

The structure of $GeO_2$ ($SiO_2$) glass can be viewed as a network of randomly distributed corner-sharing $GeO_4$ ($SiO_4$) tetrahedra. However, the $GeO_2$ tetrahedra are more distorted and the $GeO_2$ network contains significantly more 3-membered tetrahedral rings\cite{Micoulaut2006The2}. In the 50:50 $SiO_2–GeO_2$ glass, an Oxygen-17 NMR study has shown that the three species of bridging oxygen of Si-O-Si, Si-O-Ge and Ge-O-Ge appear in a ratio of about 27$\%$, 46$\%$ and 27$\%$, respectively, favouring the random mixture of the network\cite{Du2007GermanosilicateNMR}. However, Raman spectroscopy shows that the mixture is not ideal in terms of the mixing of the tetrahedra: the small (3- and 4-membered) $SiO_4$ rings are lost as they convert to large mixed $GeO_4$-$SiO_4$ rings. As a result of that, the mixed network is composed of $SiO_2$-like, $GeO_2$-like and mixed $SiO_2$-$GeO_2$-like areas. Moreover, at high temperature, the large mixed rings break up into 3-membered $GeO_4$ rings and 6-membered $SiO_4$ rings rather than the original 3- and 4-membered $SiO_4$ rings\cite{Henderson2009TheStudy}.

When alkali and alkali earth ions are added, both non-bridging oxygens can be formed with Si and Ge. However, due to the larger radius of $Ge^{4+}$, $GeO_6$ octahedra can also form, as in many crystalline germanates, depending the concentration of the modifier\cite{Du2007GermanosilicateNMR,Nishi1996StrontiumSrGe4O9}. On the thin film of $Si_{0.48}Ge_{0.52}O_2$, we did not observe the formation of $SrGe_{3.3}O_{5.6}$, in which some of Ge is likely to be 6-coordinated, as in the thin film of $GeO_2$\cite{Zhou2021CrystallizationCrystals}. This suggests that Si may prevent Ge to form high coordination compound by forming the Ge-O-Si bonds instead. On the other hand, since the mixing of the tetrahedra is not ideal due to the size mismatch, it is likely that the Sr acts differently in different areas: In a $GeO_2$-like region, since it cannot form a germanate and the local melting point may be lowered, the material starts evaporating forming the observed pits; while in a $SiO_2$-like region, Sr is more stable. As with crystallization, the evaporation process is also determined by nucleation of the new phase: the $GeO_2$-like regions initially evaporated serve as nuclei of the vapor phase and the pits expand. Fig.~\ref{fig:STO} and Fig.~S23 (see supplementary material), show an evaporated area clearly exceeding 50\% of the total area, suggesting that with the breaking of the Ge-O-Si bonds and the leaving of Ge (Ge-O), the remaining O-Si (Si) with dangling bonds will also become unstable and evaporate. In the competition between crystallization (attachment) kinetics and evaporation (detachment) kinetics, the latter wins in the film shown in Fig.~\ref{fig:STO}. Beside $Si_{0.48}Ge_{0.52}O_2$, we have also deposited thin films of $SiO_2$ on STO, and neither crystallization nor evaporation was observed with an annealing temperature of 1000 $^{\circ}C$, which proves that the $SiO_2$ network is more rigid than $GeO_2$.

\section{\label{sec:level1}Conclusions}

We have deposited amorphous thin films of $Si_xGe_{1-x}O_2$ by PLD. We have found non-stoichiometric material transfer between the target and the substrate for targets made from $SiO_2$ (quartz phase) and $GeO_2$ (quartz phase) powders, with $SiO_2$ being more resistant to the laser ablation. We solved this problem by replacing the $SiO_2$ powders with $SiO$ powders, which crystallize into cristobalite phase during the synthesis, and offer better laser absorption than the quartz phase.   

We have successfully crystallized $Si_xGe_{1-x}O_2$ with x$\leq$ 0.75 into quartz phase on quartz ($SiO_2$) substrates, which provides highly anisotropic strain conditions, achieving a higher solubility ratio than possible in the bulk crystals. These thin films crystallize at the substrate surface and grow towards the film free surface. Due to the small lattice mismatch between film and substrate, the films grow homo-epitaxially, with the same orientation of the substrate. All of these thin films are compressively strained, with the lattice in the out-of-plane direction being larger than the bulk value. With the increase of x, the mismatch between the film and the substrate is reduced. When the Si content increases to x= 0.48, the in-plane lattice is fully strained only along the direction with low-mismatch ($[1\Bar{2}10]$). When the Si content increases beyond x= 0.75, the film is fully strained. On the other hand, with the same growth procedure, the $SiO_2$ thin film shows no sign of crystallization. This indicates that by replacing Si with Ge in the network, the flexibility of the network can be enhanced, facilitating the crystallization process. 

With a view to extend the type of piezoelectric devices that can be foreseen, growth on non-piezoelectric substrates was also attempted. For $Si_xGe_{1-x}O_2$ on $Al_2O_3$ substrates, only compositions with x$\leq$ 0.16 show signs of crystallization. We believe this can be attributed to the additional nucleation barrier and worse molecular adhesion on sapphire, compared to quartz. In these thin films, the nucleation starts at the top film surface and then expand laterally, forming crystals with sizes of tens of $\mu$m. 

For thin films deposited on $SrTiO_3$, $SiO_2$ still shows no crystallization with annealing temperatures of 900 $^{\circ}C$. On the contrary, for the $GeO_2$ composition, crystallization of quartz is possible with a lower annealing temperature of 800 $^{\circ}C$ and a layer of $SrGe_{3.3}O_{5.6}$ is formed at the interface with the substrate, prior to the nucleation of quartz. With Si:Ge ratio of about 1, local evaporation and crystallization are observed with an annealing temperature of 800 $^{\circ}C$. Our findings support the idea that in a $Si_xGe_{1-x}O_2$ glass, the mixing of the $SiO_4$ and $GeO_4$ tetrahedra is not ideal due to size mismatch. At 800 $^{\circ}C$, evaporation seems to starts at the $GeO_2$-like regions, where the network is the least connected, due to the bonding to Sr. The loss of material destabilizes the surroundings, leading to further evaporation, and formation of pits. 

To conclude, the very different nucleation barriers of the two components of the $Si_xGe_{1-x}O_2$ solid-solution makes the crystallization of the these materials greatly challenging. However, it is possible to partially overcome this by using quartz substrates, on which successful growth of oriented $Si_xGe_{1-x}O_2$ single crystals have been achieved for Ge-rich compositions. These films are promising for future SAW devices where acoustic waves propagate at the surface of the film with a penetration depth comparable to the wavelength of the acoustic wave (hundreds of nm).   

\section*{SUPPLEMENTARY MATERIAL}
See supplementary material for more information on EDS analysis of the thin films and the targets, optical images of the thin films and the details of the lattice parameters in Fig.~2 and Fig.~8.

\begin{acknowledgments}
The authors are grateful to Hanna Postma and Rian Lenting for their contribution in the synthesis and analysis of the PLD targets; Kim van Adrichem for the high temperature XRD measurements; Ir.Jacob Baas for the technical support and to Adrian Carretero-Genevrier, Kit de Hond, Gertjan Koster and Guus Rijnders for useful discussions.  
This work is part of the research programme TOP-PUNT grant with project number 718.016002, which is financed by the Dutch Research Council (NWO).
\end{acknowledgments}

\section*{Data Availability}
The data that support the findings of this study are available from the corresponding author upon reasonable request.

\bibliography{main.bib}

\end{document}